\documentclass[backref, twocolumn, breaklinks, colorlinks]{aastex62}   
\hypersetup{linkcolor=blue,citecolor=blue,filecolor=cyan,urlcolor=magenta}

\usepackage{enumitem}
\usepackage{mathtools}
\usepackage{graphicx}
\usepackage{amssymb}
\usepackage{amsmath}
\usepackage{ulem}
\usepackage{epstopdf}
\usepackage{color}

\begin{document}

\title{Repeating Fast Radio Bursts from Magnetars with Low Magnetospheric Twist}

\author{Zorawar Wadiasingh}
\correspondingauthor{Zorawar Wadiasingh}
\email{zwadiasingh@gmail.com}
\affil{Astrophysics Science Division, NASA Goddard Space Flight Center, Greenbelt, Maryland, 20771, USA}
\affil{Universities Space Research Association (USRA) Columbia, MD 21046, USA}
\affil{Centre for Space Research, North-West University, Potchefstroom, South Africa }

\author{Andrey Timokhin}
\affil{Astrophysics Science Division, NASA Goddard Space Flight Center, Greenbelt, Maryland, 20771, USA}
\affil{University of Maryland, College Park (UMDCP/CRESSTII), College Park, MD 20742, USA}
\affil{Department of Physics, The George Washington University, 725 21st St. NW, Washington, DC 20052, USA}

\begin{abstract}
  We analyze the statistics of pulse arrival times in fast radio burst (FRB)~121102 and demonstrate
  that they are remarkably similar to statistics of magnetar high-energy short
  bursts. Motivated by this correspondence, we propose that repeating FRBs are
  generated during short bursts in the closed field line zone of magnetar
  magnetospheres via a pulsar-like emission mechanism.  Crustal slippage events
  dislocate field line foot points, initiating intense particle acceleration and
  pair production, giving rise to coherent radio emission similar to that
  generated near pulsar polar caps. We argue that the energetics of FRB~121102
  can be readily accounted for if the efficiency of the conversion of Poynting flux into
  coherent radio emission is $\sim10^{-4}-10^{-2}$, values consistent with
  empirical efficiencies of radio emission in pulsars and radio-loud
  magnetars. Such a mechanism could operate only in magnetars with preexisting
  low twist of the magnetosphere, so that the charge density in the closed zone
  is initially insufficient to screen the electric field provoked by the
  wiggling of magnetic field lines and is low enough to let $\sim 1$ GHz radio emission
  escape the magnetosphere, which can explain the absence of FRBs from known
  magnetars. The pair cascades crowd the closed flux tubes with plasma,
  screening the accelerating electric field, thus limiting the radio pulse
  duration to $\sim1$ ms. Within the framework of our model, the current dataset of the
  polarization angle variation in FRB~121102 suggests a magnetic obliquity
  $\alpha\lesssim40^\circ$ and viewing angle $\zeta$ with respect to the spin
  axis $\alpha<\zeta<180^\circ-\alpha$.
\end{abstract} 

\section{Introduction}

Fast radio bursts (FRBs) are curious phenomena with short $\sim 1-10$ ms observed durations, extraordinary dispersion measures, and high brightness temperatures. Repeating FRBs\footnote{also see \cite{1980ApJ...236L.109L}} \citep{2016Natur.531..202S,2019Natur.566..235C} suggest that at least some subset of FRBs originate from nondestructive events. The repeater FRB 121102 is hitherto the most well-studied with an accurate localization and distance of $d_L \sim 1$ Gpc \citep{2017ApJ...834L...7T} implying isotropic-equivalent burst energies ${\cal E}_{\rm iso} \lesssim 10^{40}$ erg. {For reviews, see \cite{2018PrPNP.103....1K,2018arXiv181005836P,2019A&ARv..27....4P}.}

Isolated neutron stars (NSs), particularly magnetars, have been suggested as a progenitor for FRB 121102 owing to the energetics of FRBs, high magnetic fields of NSs, and flaring activity of magnetars \citep{2010vaoa.conf..129P,2013arXiv1307.4924P,2014MNRAS.442L...9L,2016ApJ...826..226K,2017ApJ...843L..26B,2017ApJ...838L..13L,2018ApJ...852..140W,2019arXiv190103260L,2019MNRAS.485.4091M}, with the radio emission site being within or external to the magnetosphere. 

In this work, we advocate the view that the radio emission originates within the closed field line zone of the magnetosphere via a pulsar-like coherent emission mechanism. We motivate our model by demonstrating that FRB 121102's burst statistics bear striking similarity to short recurrent high-energy bursts of magnetars, which are a generally recognized as distinct phenomena from giant flares. Quasiperiodic oscillations, associated with crustal magnetoelastic torsional oscillations, have been reported for magnetar short bursts and therefore suggest a low-altitude crustal NS quake connection to this phenomena \citep{2014ApJ...795..114H,2014ApJ...787..128H}. 

NSs, including magnetars, are known to generate coherent radio emission. The generation of relativistic electron/positron pairs is generally accepted to be a necessary condition for operation of coherent radio emission in magnetospheres of NSs. In the galactic magnetar population, high-energy burst activity alters the current system in the magnetosphere and is associated both with the suppression \citep{2017ApJ...849L..20A} and activation \citep[e.g.,][]{2006Natur.442..892C,2018ApJ...856..180C} of coherent radio emission, presumably by altering electric fields and charge loading within the magnetosphere which regulates pair production along open magnetic field lines. 

Persistent nonthermal soft and hard X-ray emission in known magnetars of our galaxy is thought to arise via particle acceleration along closed field lines from slow dissipation of large-scale twists in a nonpotential magnetosphere with high plasma density \citep[e.g.,][]{2002ApJ...574..332T,2007Ap&SS.308..109B,2007ApJ...657..967B,2013ApJ...762...13B}. For such large field twists, the transient current density is readily satisfied for any crustal dislocations imparted on magnetic foot points (FPs) in NS crust deformations. As we show in this work, below a critical value of the field twist, this charge-abundance condition is not met and large transient electric fields necessary for avalanche pair production and operation of FRBs may result. In our model, the putative driver for FRBs is identical to short bursts in galactic magnetars, namely NS crust slippages, but differentiated by the qualitative nature of the dissipation and emission set by the state of the magnetosphere. 

 In \S\ref{obsmot}, we detail the observational motivation for our magnetar model from the polarization and burst statistics of FRB 121102. In \S\ref{model}, we describe our model, its self-consistency, and potential observational discriminants. A summary follows in \S\ref{discussion}.

\section{Phenomenological Motivation from FRB 121102}
\label{obsmot}

\subsection{Lognormality of Bursts and Power-law Fluence Distributions}
\label{burststats}

\defcitealias{2018ApJ...866..149Z}{Z18}

Seemingly random recurrent high-energy bursts from galactic magnetars are common and a defining trait, with a broad energy range {$\sim 10^{36}-10^{42}$} erg. They are superficially distinct from FRBs, with $T_{90}$ durations $0.01-1$ s, i.e. $\sim 10$ to $10^3$ times longer than FRB pulses. Yet, although the radiative processes are dissimilar, the underlying driver may be identical by virtue of the occurrence and fluence distributions. In contrast to giant flares, there is evidence for confinement of plasma (in closed zones) rather than outflows in short bursts. The fluence range implies $\ll 10^{-3}$ fractional depletion of the $\sim 10^{46}-10^{48}$ erg magnetic reservoir per burst. The high-energy spectrum of short bursts is quasithermal and may be described by a two-blackbody model \citep[e.g.,][]{2008ApJ...685.1114I,2012ApJ...749..122V,2012ApJ...756...54L,2014ApJ...785...52Y,2015ApJS..218...11C}. The two-blackbody model indicates temperatures $T_{\rm cool} \sim 3-5$ keV and $T_{\rm hot} \sim 10-50$ keV, with inferred cool and hot emission areas $[(0.3-1)R_{*}]^2$ and $[(0.03-0.1) R_{*}]^2$ ($R_*$ the radius of the NS), respectively, with similar flux in both components. The hotter component is indicative of compactness and generally interpreted as arising from hot spots localized near magnetic field line FPs. Indeed, changes in the soft X-ray pulse profiles and surface heating are ubiquitous during such short burst episodes.

\begin{figure}[t]
\centering
\includegraphics[width=0.45\textwidth]{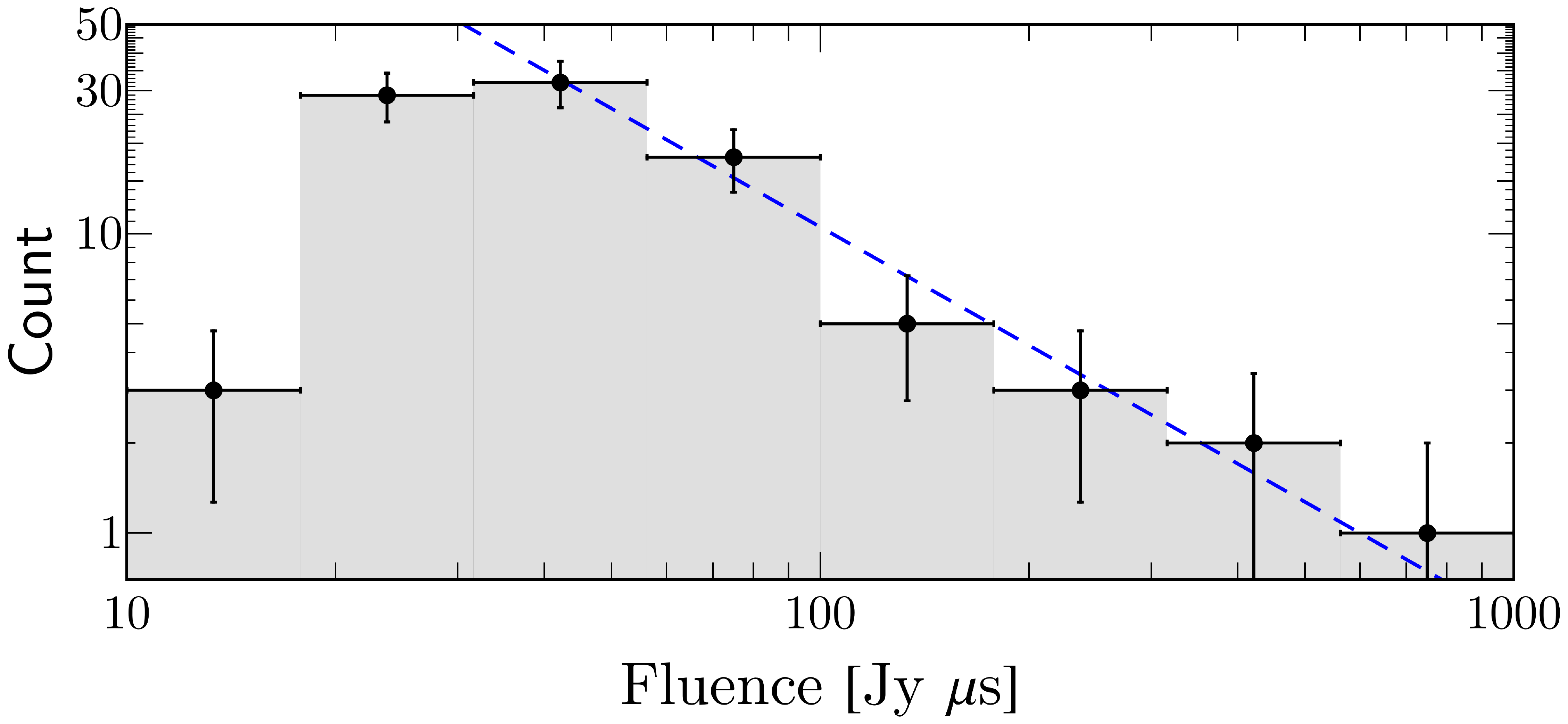}
\caption{Histogram of fluences of bursts from FRB 121102 (with Poisson uncertainties) from \citetalias{2018ApJ...866..149Z}. The blue dash line depicts a power law $N \sim 4470{\cal F}^{-\gamma+1}$ with index of $\gamma = 2.3$, for ${\cal F} \gtrsim 30$ Jy $\mu$s.}
\label{fluencehist}
\end{figure}

The phenomenology of short bursts is worth noting for comparison with FRB 121102. Firstly, magnetar short bursts are episodic: intense activity with hundreds of bursts in hours may be followed by inactivity of months/years. Such episodic behavior is generally predicted by recent magnetothermal models of magnetar crustal stresses \citep[e.g.,][]{2011ApJ...727L..51P,2013MNRAS.434..123V,2015MNRAS.449.2047L}. 
Second, the waiting time distribution of bursts within episodes is lognormal \citep{1994A&A...288L..49H,1999ApJ...526L..93G,2000ApJ...532L.121G,2004ApJ...607..959G,2010A&A...510A..77S}, typically with mean $\sim 100$ s and width $\sim 1$ dex . The waiting time of bursts may be correlated with arrival time, but generally no robust correlation exists for fluence with waiting or arrival time \citep{1996Natur.382..518C}. {Moreover, evidence for a spin-phase dependence of short bursts is generally weak \citep[e.g.,][]{2015ApJS..218...11C}; the duty cycle of bursts over a spin period can be broad and weakly varying over a known rotational ephemeris. If the bursts are intrinsically beamed, strong evidence for phase dependence is not expected to emerge without significantly larger samples of bursts \citep{2018MNRAS.476.1271E} owing to the large separation of timescales between the short burst durations and long spin period. The inverse problem of establishing periodicity from burst arrivals would clearly be challenging for a limited collection of bursts. } Thirdly, the differential distribution of fluences ${\cal F}$ can be described by a power law $dN/d{\cal F} \propto {\cal F}^{-\gamma}$ with $\gamma \sim 1.4 - 2.0$ for a multitude of burst episodes \citep[e.g.,][and references therein]{2015RPPh...78k6901T}.

To date, \cite{2018ApJ...866..149Z} (Z18) report the largest public sample of FRB 121102 bursts observed on August 26, 2017 at $4-8$ GHz at the GBT. \citetalias{2018ApJ...866..149Z} report no evidence of periodicity in burst arrival times. In that sample of 93 bursts spanning five hours of continuously telescope coverage, the fluence of bursts varies ${\cal F} \in [13, 606 ]$ Jy $\mu$s with standard uncertainties of $\delta {\cal F} \sim 10-15$ Jy $\mu$s, with an instrumental threshold of $\sim 10-30$ Jy $\mu$s. Assuming a flat spectral index over bandwidth $\Delta W$, this implies isotropic-equivalent fluences
\begin{equation}
{\cal E}_{\rm iso} \lesssim 3 \times 10^{39} \left( \frac{{\cal F}}{600 \, \rm Jy \, \mu s } \right) \left( \frac{\Delta W}{4 \rm \, GHz } \right) \left( \frac{d_L}{1 \rm \, Gpc } \right)^2 \quad \rm erg.
\end{equation}

\begin{figure}[t]
\centering
\includegraphics[width=0.45\textwidth]{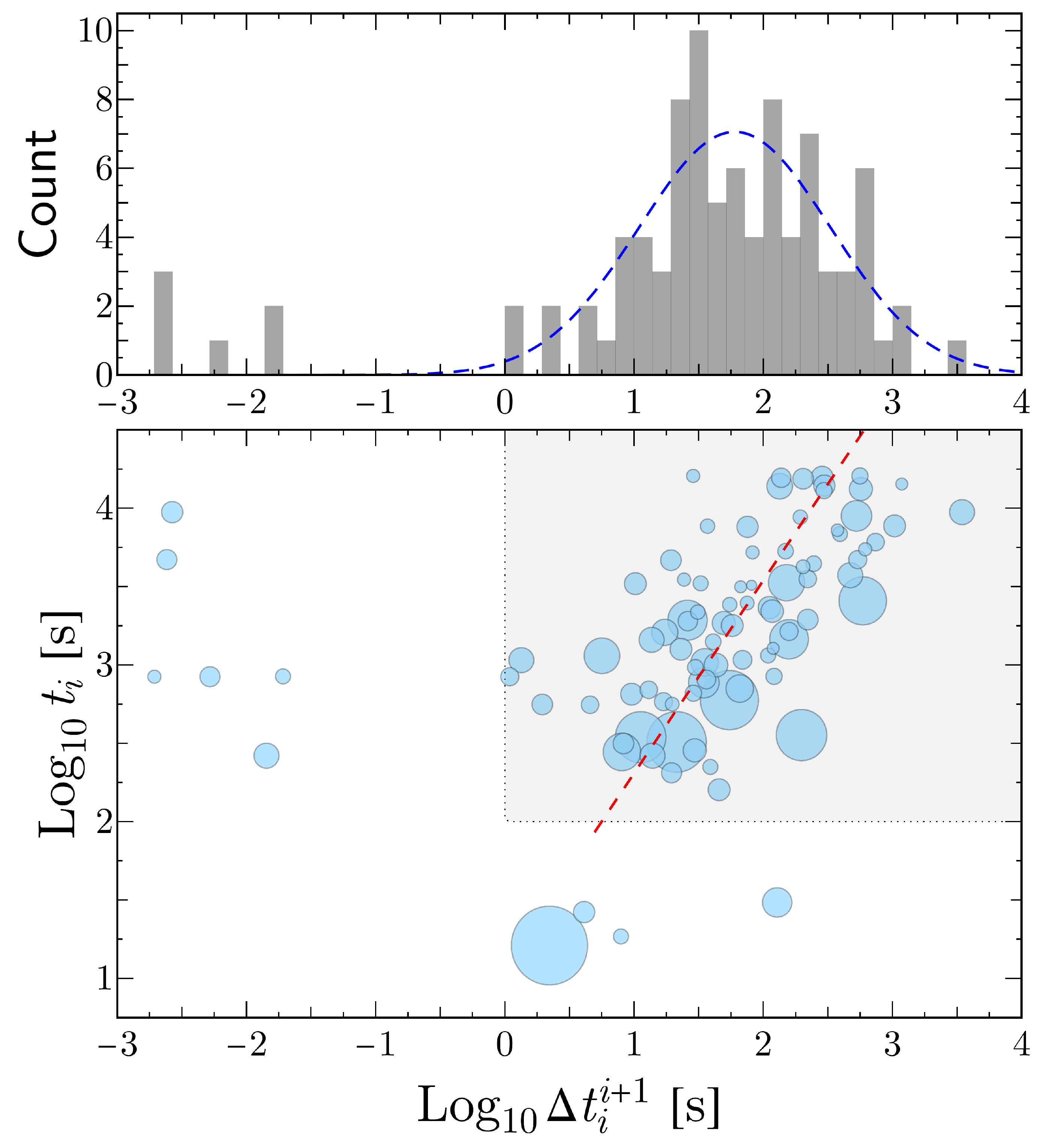}
\caption{{\bf{(Bottom)}} Next burst waiting time $\Delta t^{i+1}_{i}$ versus arrival time $t_i$ for FRB 121102 \citepalias{2018ApJ...866..149Z}. The area of circles is proportional to fluence. For the boxed cluster, a power-law $\Delta t^{i+1}_{i} = 0.13\, t_i^{0.81}$ is represented by the dashed red line. {\bf{(Top)}} Histogram of $\log_{10} \Delta t^{i+1}_{i}$. The dashed blue curve illustrates a lognormal distribution with mean $60$ s and width $0.74$ dex.}
\label{deltaTOA}
\end{figure}

In Figure~\ref{fluencehist}, we display a coarsely binned (bin width~$\gg \delta{\cal F}$) histogram of fluences for the \citetalias{2018ApJ...866..149Z} sample. Kolmogorov-Smirnov (KS) and Anderson-Darling (AD) tests both strongly reject a purely exponential distribution of fluences above  ${\cal F} > 30$ Jy $\mu$s  ($p \sim 10^{-7}-10^{-6}\ll 0.01$) while they do not reject a power-law distribution. Above ${\cal F} \gtrsim 30$ Jy $\mu$s, we obtain a binned Poissonian maximum likelihood fit, $dN/d{\cal F} \propto {\cal F}^{-\gamma}$  with $\gamma \sim 2.3 \pm 0.2$. Via Monte Carlo exploration, we notice that if events are drawn from this power-law distribution, the paucity of low fluence events below $\sim 30$ Jy $\mu$s is consistent with a toy model with an instrumental threshold of $\sim 10 $ Jy $\mu$s and $\delta {\cal F} \sim 10$ Jy $\mu$s. The index $\gamma \sim 2.3$ is somewhat steeper than that for magnetar short bursts, although the limited statistics warrant a larger sample to test for any Weibull distribution-like curvature/cutoff to the power law. The steeper index of fluences may be regulated by the efficiency of the emission process or a propagation effect. 

 In Figure~\ref{deltaTOA}, the next-burst waiting time ($\equiv \Delta t^{i+1}_{i} = t_{i+1} - t_i$) from \citetalias{2018ApJ...866..149Z} is depicted {\citep[see also][]{2019MNRAS.487..491K}}. There are three salient features worth highlighting. For the bulk of events where $\Delta t^{i+1}_{i} > 1 s$, we find that the waiting time distribution of bursts is consistent with a lognormal distribution {of mean $\approx 60$ s ($50$ s in the source frame)}. Second, the waiting time of bursts is correlated with arrival time as in magnetar short bursts \citep[e.g.,][]{1996Natur.382..518C,2004ApJ...607..959G}. A least squares analysis for the cluster ($\Delta t^{i+1}_{i} > 1$ s and $t_i > 100$ s) in Figure~\ref{deltaTOA} yields $\Delta t^{i+1}_{i} \approx 0.13 ^{+0.18}_{-0.08} \, t_i^{0.81 \pm 0.11}$. We note the surprising consistency of this phenomenology to that reported for 1E 2259+586 \citep[cf. Fig. 10 of][]{2004ApJ...607..959G}. Finally, there is no significant dependence of burst fluence with event and waiting time in log-log (coefficient of determination $r^2 \approx 0.1$ and $0.02$, respectively) for the cluster highlighted in Figure~\ref{deltaTOA}. Thus, for the bulk of events, the statistics of FRB events in FRB 121102 bear \emph{striking similarity to magnetar short bursts.}
 
 \cite{2019arXiv190302249G} also remark on lognormality of waiting times in a collection of FRB 121102 Arecibo bursts, but that sample is insufficient to establish the power-law relation as in Figure~\ref{deltaTOA}.
 
 \begin{figure}[t]
\centering
\includegraphics[width=0.45\textwidth]{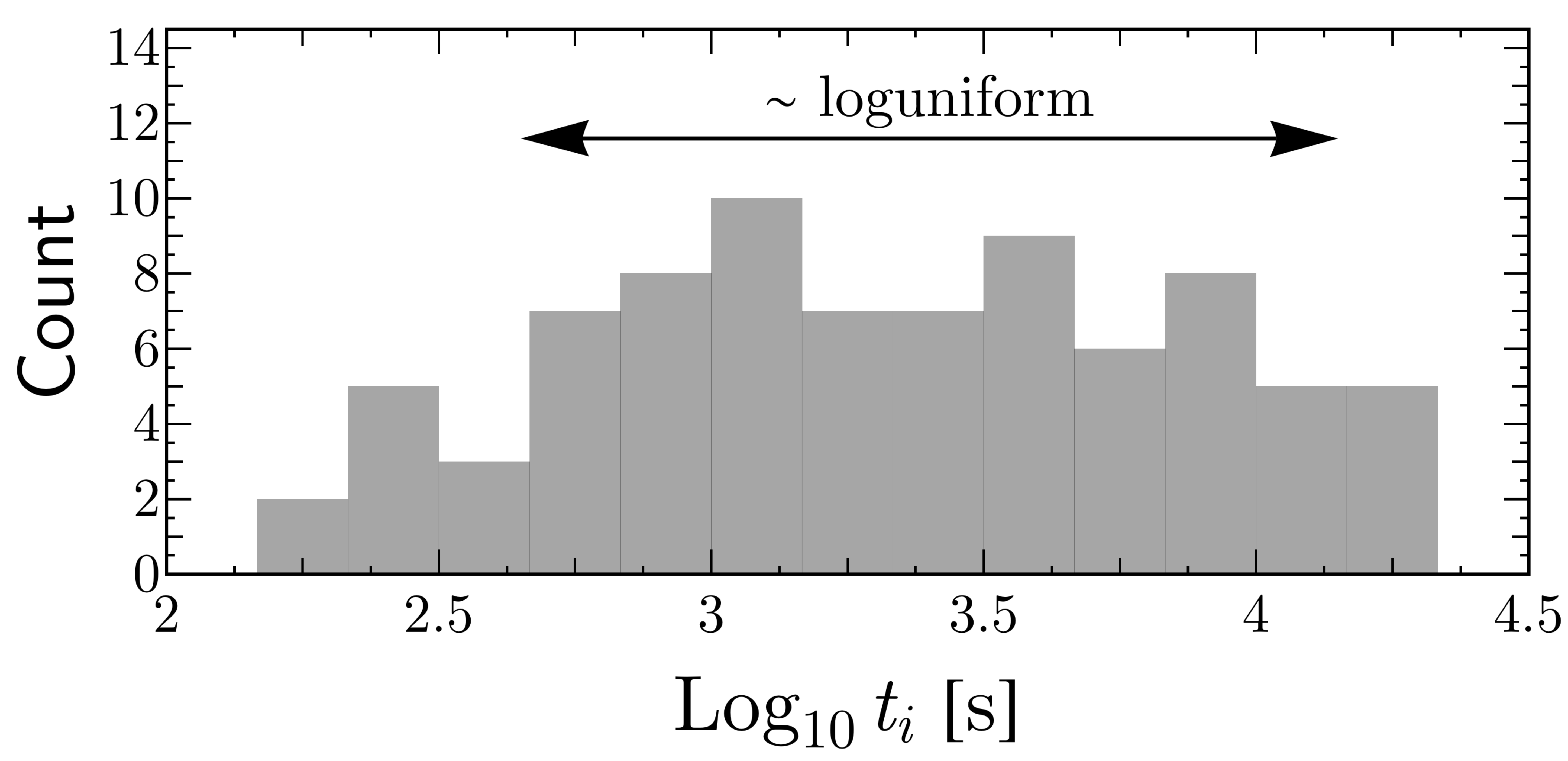}
\caption{Coarse-grained loguniformity of arrival times for the cluster of events highlighted in the bottom panel of Figure~\ref{deltaTOA}. }
\label{loguni}
\end{figure}

Figure~\ref{deltaTOA}, {which was constructed as an analog to Fig.~10 of \cite{2004ApJ...607..959G}}, may be understood as follows. The arrival times exhibit approximately a loguniform distribution {for the event density (number of events in time interval $dt$) $N(t) dt \propto dt/t = d\log t$} for the boxed region of Figure~\ref{deltaTOA} (this nonstationary Poissonian character was noted by \citetalias{2018ApJ...866..149Z} but not its form). {The coarse-grained loguniform nature of arrival times is depicted in Figure~\ref{loguni}, which is a reduction of the highlighted region in the bottom panel of Figure~\ref{deltaTOA} onto the vertical axis. KS/AD tests do not reject this description. Formally, the arrival times can then be regarded as an ``order statistic" of random variables from a continuous loguniform distribution. The log character implies scale invariance (suggestive of multiplicative physical processes with {\textit{memory}}) and signifies that the relative ratios of arrival times rather than offset to an arbitrarily assumed zero time is what is relevant. The ``power-law relation" in Figure~\ref{deltaTOA}, which simply captures the gross trend of increasing waiting time with arrival time, is an empirical construct whose origin can be traced to the arrival time loguniformity. The fit exponent of this relation is contingent of the dynamic range of the timescales during a burst storm, and would clearly be very poorly constrained if the dynamic range of timescales were small. Conversely, the fit exponent approaches unity (from below) for a large dynamic range of timescales where loguniformity is realized, i.e. when the indicated range with arrows in Figure~\ref{loguni} is more extensive. 

Under the assumption that the arrival times are an order statistic of random variables from a loguniform distribution, the waiting time distribution is humped and asymptotes to loguniformity for infinite episode duration. The lognormal distribution (parabola in log-log) is a low-order nontrivial approximation for a humped distribution at the peak. The number of events and truncated sampling (i.e. the dynamic range of timescales in Figure~\ref{loguni}) of the $\propto 1/t$ event density governs the mean and width of the distribution -- this phenomenology may be regulated by physics that sets the characteristic duration of burst episodes to hours/days and the number of bursts in the tens or hundreds. Hence, the similarity of FRB 121102 and magnetar short burst phenomenology is fundamentally linked by the $1/t$ (or loguniformity) event density underpinning burst triggering during episodes and by the comparable characteristic lifespan of such burst storms. This fundamental similarity is one aspect which motivates our model in \S\ref{model}.}

The six short waiting time events in Figure~\ref{deltaTOA}, if not spurious, may reflect double-peaked events below instrumental threshold -- such short waiting time events are also encountered in studies of magnetar short bursts. {As will become apparent in due course, we ascribe a different physical origin for these events in our model than events which follow the gross trend in Figure~\ref{deltaTOA}. }

\subsection{PA Stability During and Between Bursts}
\label{obsPA}

FRB 121102 exhibits $\sim 100\%$ linear polarization in its pulses. Of the 16 Arecibo/GBT bursts reported in \cite{2018Natur.553..182M}, 13 have measured polarization angles (PAs), and exhibit a sample mean of $\langle \rm PA \rangle \sim 63^\circ$ and standard deviation $\sigma_{\rm PA} \sim 8^\circ$, i.e. a relatively narrow range of PAs. For this sample, KS and AD tests disfavor ($p<0.05$) a bracketed uniform distribution $\rm PA \in [\langle PA \rangle - \chi, \langle PA \rangle + \chi ]$ for range $\chi \gtrsim 40^\circ$, suggesting total chaos ($\chi = 90^\circ$) is improbable. Similarly, for the 13 PA measurements (out of 21 GBT bursts) \cite{2018ApJ...863....2G} report, $\langle \rm PA \rangle \sim 77^\circ$ with standard deviation $\sigma_{\rm PA} \sim 7^\circ$ with $\chi \gtrsim 30^\circ$ disfavored by KS/AD tests. The modestly different sample means $\langle PA \rangle$ of the two datasets may reflect the dissimilar sampling cadences. 

Both \cite{2018Natur.553..182M} and \cite{2018ApJ...863....2G} reported that during bursts, the PA was fixed to within $\sim 5^\circ-10^\circ$. As noted by \cite{2018Natur.553..182M}, this is suggestive of an emission process where the observed burst duration (apart from scattering broadening) is intrinsic rather than a geometric effect of an observer intercepting a sweeping beam from a polar cap\footnote{In radio pulsars where radio emission arises from the polar cap open zone, the PA can sweep significantly ($\gg 5^\circ$) during a single pulse \citep[e.g.,][]{2001ApJ...553..341E}. This is generically true when the pulsar magnetic obliquity is appreciably nonzero.}. Modulo viewing geometry influences, this is a natural consequence if the phase width of the beam $\delta \phi$ is wide in comparison to the ratio of the intrinsic burst duration $\tau$ to the period of the rotator $P$, and $\delta \phi \gg \tau/P$ where for slow rotators like magnetars $\tau/P \ll 1$. Indeed, the high rate of bursts, the null-correlation with rotational phase, and statistical similarity to magnetar short bursts suggest FRB 121102's beaming cones are broad. Then, geometry is likely the driver of PA variation between bursts.
  
\section{The Charge-starved Magnetar Model}
\label{model}

The striking similarity between the fluence and recurrence rate phenomenology of
magnetar short bursts and FRB 121102 pulses motivates us to consider a model
where repeating FRBs are generated in magnetospheres of (some) magnetars
experiencing short bursts. Additional arguments in favor of such an explanation
would be the fact that highly magnetized NSs -- pulsars and some magnetars -- do
exhibit coherent radio
emission and the high magnetic fields of NSs would help to explain the high
polarization seen in the pulses of FRB 121102 \citep[some other FRBs also
exhibit high linear polarization
e.g.,][]{2015Natur.528..523M,2017MNRAS.469.4465P,2018MNRAS.478.2046C} -- such
linear polarization is suggestive of either generation in or natural eigenmode
propagation within strong and ordered magnetic fields
\citep[e.g.,][]{2017RvMPP...1....5M}. 

In order for such a model to be viable it must at least account for (i) the
energetics of individual radio pulses, ${\cal E}_{\rm iso} \sim10^{37}-10^{39}$
erg, (ii) their short duration, $\lesssim 1$ ms, (iii) explain why such coherent
high intensity burst-like radio emission is not seen from known magnetars during bursts or perhaps at other epochs. Here we address these items and develop
a schematic model for repeating FRBs generated by magnetars.

\begin{figure}[t]
\centering
\includegraphics[width=0.4\textwidth]{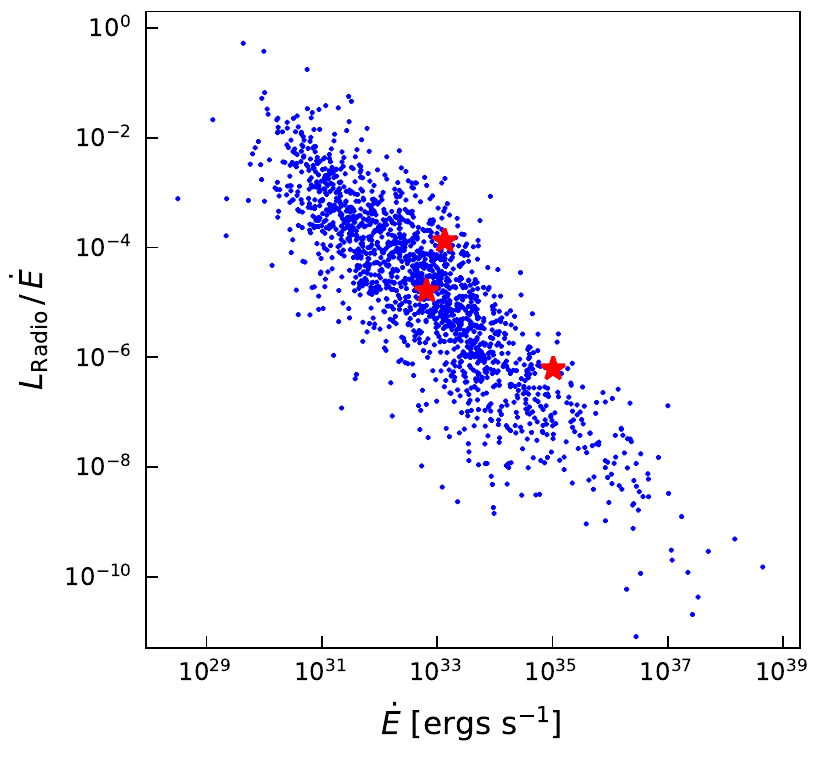}
\caption{Observed radio efficiency of pulsars (magnetars in red) at 1.4 GHz \citep[ATNF catalog,][]{2005AJ....129.1993M}.}
\label{LR}
\end{figure}

In Figure~\ref{LR} we display the ratio of radio luminosity to spindown power
for pulsars and 3 known AXPs observed to emit pulsed radio signals sporadically%
\footnote{We adopted values for radio flux at $1.4$ GHz multiplied
  by the square of the pulsar distance in units of [mJy~kpc$^{2}$] (quantity $\texttt{R\_Lum14}$
  in ATNF catalog) and calculated the luminosity as
  $L_{\mathrm{Radio}}= 9.5\times10^{16} \times 4\pi\,\texttt{R\_Lum14}\,
  \Delta\nu$~[erg s$^{-1}$] with $\Delta\nu=1$ GHz.}.
It is evident from this plot that the pulsar emission mechanism(s) could operate
with the efficiency in the range of $10^{-2}-10^{-6}$ in physical conditions
present in NS magnetospheres, converting Poynting flux into coherent radio
emission; attempts to correct for beaming empirically
\citep{2002ApJ...568..289A} may further relax this efficiency constraint,
especially for pulsars near the death band (i.e. charge starvation). The leading
model for magnetar short bursts invoke deformations of the NS crust. For
magnetoelastic deformations, a characteristic energy scale of $10^{42}-10^{43}$
erg is plausibly attainable
\citep[e.g.,][]{2001ApJ...561..980T,2011ApJ...727L..51P,2015MNRAS.449.2047L}. The
quasithermal short burst {energies up to $\sim 10^{42}$} erg are then
\textit{calorimetric} for the event energy release into the magnetosphere. For FRB 121102, a
pulsar-like emission mechanism converting $\sim 10^{-4}-10^{-2}$ of the total
energy released into the magnetosphere may comfortably account for the
energetics of radio bursts without invoking beaming;
such efficiency is, at least, not inconsistent with estimated efficiencies of
pulsar emission mechanisms shown in Figure~\ref{LR}.

Although the specifics of the emission mechanism(s) are unknown, it is
generally accepted that in most pulsars coherent radio emission is generated along open magnetic field lines at low altitudes. The critical
ingredient for this mechanism is the presence of cascade zones where particles
are accelerated to high energies in vacuum-like gaps. These particles emit high energy
$\gamma$-rays which give rise to copious electron/positron pair cascades via magnetic pair
production. In the process of such highly nonstationary plasma outflow,
coherent radio emission is putatively generated. The basis for the existence of particle
acceleration zones is the repeated depletion in some magnetospheric locales of charge carriers, which are transported into the pulsar wind, 
and the resulting inability of these regions to sustain current densities demanded by the magnetosphere. As the particle number density drops below the value
necessary to support the current and charge densities required by the global
magnetospheric configuration, a quasivacuum gap with high electric field
appears \citep[e.g.,][]{2010MNRAS.408.2092T,2013MNRAS.429...20T}. The
characteristic charge density $\rho_{\mathrm{GJ}}$ needed to screen the
accelerating electric field is the Goldreich-Julian charge density
\citep{1969ApJ...157..869G}, the critical current density $j_{\mathrm{m}}$
varies over the polar cap, but for most pulsars is in the range
$| \boldsymbol{j_{\mathrm{m}}}|\lesssim (1-2)\,\rho_{\mathrm{GJ}}\, c$. In the closed field line zone,
plasma is trapped. In a rotation-powered pulsar, no currents flow along closed field lines and plasma there, once
generated, may persistent subject only to slow
diffusion-like processes, thus hindering the formation of acceleration zones and the
generation of coherent emission.  In the de-facto standard magnetar model
\citep{2002ApJ...574..332T,2007ApJ...657..967B} the magnetic field has a global
twist which demands persistent ``simmering'' pair creation to support current
flow along closed magnetic flux tubes. Plasma does flow along those
field lines but is constantly replenished by low-intensity pair
formation.

In a magnetar short burst, $\sim1\%$  of the
NS surface area participates in the energy release \citep[$T_{\rm hot}$ component, e.g.,][]{2008ApJ...685.1114I,2012ApJ...749..122V,2012ApJ...756...54L}, which is much larger than the
polar cap area\footnote{The polar cap is only a small fraction of the total NS surface area, $\pi r_{\mathrm{pc}}^{2}/(4\pi R_{*}^{2}) \simeq 5\times 10^{-5}\,P^{-1}$ with $P$ in seconds.}.
Therefore, in magnetar crustal slips, most of the energy release will involve the closed magnetic
flux tubes.  During such events, an electric field will be
generated owing to dislocation of magnetic FPs.  If the plasma
density proximate to this active region is sufficient to screen this electric
field, no pulsar-like emission mechanism may operate. Indeed, if charges are abundant, the characteristic size of regions with unscreened electric fields is of the order of the Debye length, and particles will not be accelerated to energies high enough to initiate strong cascades, e.g. \cite{2007ApJ...657..967B}. However, if the plasma density in the closed flux
tube is below the Goldreich-Julian density associated with the burst event%
\footnote{i.e. the charge density necessary to screen $E$ precipitated by magnetic FP dislocations.},
e.g. due to low initial field twist, then magnetospheric
regions linked to this active area may become charge starved,
what will lead to intense particle acceleration, pair creation, and generation
of coherent radio emission via a putative pulsar-like mechanism. In this case, a larger
area would emit more coherent radio emission than that in pulsars/magnetars, where the
emission is limited by the open zone.

Let us now estimate the critical twist of closed dipolar magnetic field lines which
would allow the operation of a pulsar-like mechanism.  The current density needed to
support a persistent field line twist is \citep{2007ApJ...657..967B}
\begin{equation}
  \label{eq:j_twist}
  j_{\textrm{twist}}=\frac{c}{4\pi} \left| \boldsymbol{\nabla}\times\boldsymbol{B}  \right|\sim
  \frac{c}{4\pi}\frac{B}{R_{*}}\,\sin^{2}\theta_0 \,\Delta\phi\,
\end{equation}
where $B$ is the local magnetic field, $\theta_0$ is the FP colatitude and $\Delta\phi$ is the twist
angle. The corresponding charge density is of order
$\rho_{\textrm{twist}}\sim{}j_{\textrm{twist}}/c$.

Wiggling of magnetic FPs with the speed $v$ will result in an
electric field
\begin{equation}
  \label{eq:E}
  E\sim{}\frac{v}{c} B \sim \frac{2\pi{}\nu\xi}{c}B
\end{equation}
where $\nu$ and $\xi$ are the frequency and amplitude of oscillations, respectively.  The requisite
charge density to screen the accelerating electric field $E$ in the active
region provoked by wiggling of magnetic FPs may be estimated as
\begin{equation}
  \label{eq:pho_GJ_osc}
  \rho_{\textrm{burst}}\sim\frac{1}{4\pi}\frac{E}{\lambda}\sim
  \frac{1}{2}\frac{\xi}{\lambda}\frac{\nu}{c}B
\end{equation}
where $\lambda$ is the characteristic wavelength of oscillations. When
\begin{equation}
  \label{eq:rho_inequality}
  \rho_{\textrm{burst}}> \max \{ \rho_{\textrm{twist}}, \rho_{\rm GJ} \}\,
\end{equation}
the charge density is insufficient to screen $E$ prompted by the NS crust motion
and the resulting charge starvation would give rise to intense particle
acceleration and, according our assumptions, an FRB.
The charge density due to
the twist is larger than the corotational Goldreich-Julian one
$\rho_{\rm GJ} \sim B/(c P)$ provided that
$\Delta\phi~\gtrsim~4\pi R_{*}/(c P \sin^2 \theta_0) \sim 4\times 10^{-5} (P/10 {\rm \, s})^{-1}
\sin^{-2} \theta_0 $, and so we neglect $\rho_{\rm GJ}$ here onwards.
Influences of corotation on the twisted
currents are only relevant for altitudes much larger than considered here
\citep{2002ApJ...574..332T}. 

From Eqs.~\eqref{eq:j_twist}--\eqref{eq:rho_inequality}, we obtain
the limit on the preexisting local twist of magnetic field lines which allow a pulsar-like
emission mechanism to operate in the closed zone,
\begin{equation}
  \label{eq:delta_phi}
  \Delta\phi \lesssim
  \frac{2\pi{}R_{*}}{c}\nu\frac{1}{\sin^{2}\theta_{0}}\frac{\xi}{\lambda}
  \simeq 0.003\, \nu_{\mathrm{kHz}}\,\, \sigma_{-3}\,.
\end{equation}
The last step in the inequality assumes that the colatitude of magnetic field FPs $\theta_{0}\simeq15^{\circ}$ (corresponding to flux tube filling times of $\sim$1 ms --
see below); $\nu_{\mathrm{kHz}}$ is the oscillation frequency
in kHz, and the strain $\sigma\equiv\xi/\lambda$ is normalized to $10^{-3}$,
$\sigma_{-3}\equiv10^{-3}\left(\xi/\lambda\right)$, following usual assumptions
about properties of magnetar crusts
\citep[e.g.,][]{1995MNRAS.275..255T}. Eq.~(\ref{eq:delta_phi}) may be applied to a single dislocation of duration $\Delta{t}$ as well, by $\nu \rightarrow 1/\Delta{t}$. For crustal breakage events,
the strain will be larger than that for oscillations, hence, a crustal failure 
event may generate intense pair cascades in magnetars with larger initial twist. The limiting
twist Eq.~(\ref{eq:delta_phi}) for typical parameters associated with magnetar bursts is lower than
usually needed for persistent nonthermal emission in active magnetars
\citep[e.g.,][]{2007Ap&SS.308..109B,2007ApJ...657..967B,2013ApJ...762...13B} and
confirms the basic expectations of our model. Note that Eq.~(\ref{eq:delta_phi})
is independent of $B$.

The maximum potential drop which may be generated by crustal displacements is
of the order of $\Delta\Phi_{\max}\sim{}E\,\lambda$, for $E$ given by
Eq.~\eqref{eq:E}. The upper limit on the energy of primary electrons
accelerated by a parallel electric field above the active region would be
\begin{equation}
  \label{eq:Phi_max}
  \gamma_{\max}\sim
  \frac{e\Delta\Phi_{\max}}{m_{e}c^{2}}\sim
  10^{9}\,\, \nu_{\mathrm{kHz}}\, \sigma_{-3}\,\lambda_{4}^{2}\,B_{14}\,, 
\end{equation}
where $m_{e}$ and $e$ are electron mass and charge, and $\lambda_{4}\equiv{}\lambda/10^{4}$ cm -- characteristic wavelength of
oscillations/displacement are normalized to $1\%$ of the NS radius. It is evident that
for any reasonable values of parameters, in the case of charge starvation, primary
particles will achieve energies sufficient to trigger pair cascades. { In our model, particle acceleration commences at the beginning of a short burst when burst-induced photon densities are low. Then, particle Compton drag can only arise by scattering soft thermal X-ray photons from the NS surface. The acceleration rate of a primary $\dot{\gamma}_e \sim e/( m_e c) E \sim 10^{15.5} B_{14} \nu_{\mathrm{kHz}} \sigma_{-3} \lambda_4$ \, s$^{-1}$ ($B_{14}$ is the magnetic field in units of $10^{14}$ G) is much greater than even the peak ($\sim 10^{9}-10^{12}$ \,s${}^{-1}$) of the resonant Compton\footnote{In magnetars, resonant Compton scattering is the dominant energy loss mechanism for electrons at modest Lorentz factors at low altitudes.} cooling rate in a surface thermal photon bath \citep[see Figures 4--6 in][]{2011ApJ...733...61B}; hence, the primary will accelerate until curvature losses dominate at $\gamma_e \gtrsim 10^6$. The existence of persistent polar cap coherent radio emission in some magnetars \citep[e.g.,][]{2007MNRAS.377..107K} also provides strong evidence that such Compton drag is not a showstopper.}

Regardless of the actual radio emission mechanism, the radio waves ought to
decouple from the magnetosphere and escape to infinity to be observable. This is
involved, and ultimately hinges on the dielectric tensor and anisotropic plasma
dispersion relations in the strongly magnetized quantum plasma\footnote{In
  strong magnetic fields, vacuum polarization may dominate the dielectric tensor
  for wave propagation. However, its impact strongly weakens for lower energy
  photons. For radio photons, it can be shown that the characteristic pair
  number density below which vacuum birefringence dominates over plasma effects
  \citep[the vacuum resonance condition e.g.,][]{2002ApJ...566..373L}, is far
  lower than even $\rho_{\rm GJ}$ for any reasonable plasma bulk Lorentz
  factor.}. Among other factors, such as the direction of wave propagation with
respect to local $\boldsymbol{B}$, the unknown local plasma distribution
function and bulk Lorentz factor in the NS frame can influence cutoffs, with
higher transparency for larger bulk motions. Conservatively, the plasma
frequency $\nu_{e}$ (with zero bulk motion) for the plasma supporting the twist of the closed field lines sets the
characteristic wave frequency scale, below which radio emission is likely damped
or anomalous,
  \begin{align}
 \nu_{e}\sim\frac{1}{2\pi}\sqrt{\frac{4\pi{}e\rho_{\mathrm{twist}}}{m_{e}}}\sim \frac{1}{2\pi} \sqrt{\frac{e B}{m_{e}R_{*}}}\, \sin\theta_0 \,\Delta\phi^{1/2}
    \label{eq:nu_e_general}
  \end{align}
adopting Eq.~\eqref{eq:j_twist}. For the limiting twist Eq.~(\ref{eq:delta_phi}), a limit on the plasma frequency is 
  \begin{align}
    \label{eq:nu_e}
    \nu_{e}\lesssim & \frac{1}{\sqrt{2\pi}}
    \omega_{B}^{1/2}\,\,\sigma^{1/2}\,\nu_{\mathrm{osc}}^{1/2}\, \sim \nonumber \\ 
  &  17\, B_{14}^{1/2}\,\sigma_{-3}^{1/2}\,\nu_{\mathrm{osc,\, kHz}}^{1/2}\quad\mathrm{GHz}\, ,
  \end{align}
 where $\omega_B = e B/(m_e c)$. Radio waves of frequencies $\nu_{\rm em}\sim 1$ GHz {(in the source frame)} may
  escape from the low-twist magnetosphere beginning at about altitudes
  $r_{\mathrm{em}}$ where magnetic field drops below $B_{14}\lesssim(1/17)^{2}$,
  for $r_{\mathrm{em}}\gtrsim7\,R_{*} B_{0, 14}^{1/3}$, where $B_{0, 14}$ is the
  surface magnetic field (normalized to $10^{14}$ G). Dipolar magnetic flux
  tubes can extend up to maximum altitude
  $r_{\max}\simeq R_{*}\sin^{2}\theta_{0}$ -- from the limit
  $r_{\mathrm{em}}\gtrsim 7 R_*$ for $1$ GHz propagation, we obtain a
  restriction on the FP colatitude
  $\theta_{0,\,\mathrm{em}}\lesssim\arcsin(\sqrt{R_{*}/r_{\mathrm{em}}})\simeq23^{\circ}$. For
  twists smaller than the critical one, the range of magnetic field lines along
  which the emission can escape is larger, as follows from
  Eq.~\eqref{eq:nu_e_general}. {Note that Eq.~(\ref{eq:nu_e}) in some sense may be regarded as a radius-to-frequency mapping \citep[e.g.][]{1978ApJ...222.1006C}.}

  Above, we examined limits on the emission height considering the transparency of plasma generated by the persistent twist of magnetic field lines, without considering transparency of plasma that generates the radio emission. This plasma, generated in the event leading to the FRB, ought to be much denser than the
  background plasma through which the radio emission propagates, but also relativistic. Then,
  the altitude of transparency may be larger than that estimated above, and the colatitudes of magnetic field lines
 FPs may be smaller than $\theta_{0,\,\mathrm{em}}$. However,
 details of the putative pulsar-like emission mechanism are poorly understood, and the straightforward
  arguments used above might not be applicable to the emission regions above the
  active zone. Moreover, as in pulsars, field curvature is expected to play a role in transparency. Hence, the estimates for the extent of the regions from which the
  GHz radio emission can escape based on the background plasma density represent
  an upper limit on the size of those regions. 

In our scenario, the pulsar-like mechanism may operate only until the dense pair
plasma fills the closed flux tube originating in the active zone. Then, even if
the motion of field line FPs persists on longer timescales, plasma density will
remain high, and particle acceleration and the associated coherent emission will be
stifled. The time needed to supply charges for a flux tube extending up to the
maximum distance $r_{\max}$ will be of the order of $\tau\sim{} 2 r_{\max}/c$. For
the dipolar field, the FP colatitude for the flux tube which will be populated by
plasma in $\tau\lesssim 1$~ms is $\theta_{0,\,1\mathrm{ms}}\gtrsim14^{\circ}$.

The clearing of flux tubes permeated by plasma from pair cascades is not immediate. If
these field lines were twist-free, the plasma may persist a long time, subject to slow diffusion-like processes, and hinder subsequent FRBs if the same FP is dislocated. If those field loops are mildly twisted, clearing can
proceed faster as charged particles will be exhausted for supporting current
flowing along these field lines due their twist. The minimum time required for clearing
of the flux tube of length $\ell_B \sim 2 r_{\rm max}$ of pair plasma can be estimated as
$\sim{}\ell_B\,\kappa\rho_{\textrm{burst}}/j_{\textrm{twist}}$, where $\kappa$ is the
multiplicity of the pair cascade. For the case of
near-threshold twists, when $\rho_{\textrm{burst}}\sim{}\rho_{\textrm{twist}}$,
the minimum interval between bursts would be $\kappa$ times longer than the burst
duration. For expected values $\kappa\sim10^{2}-10^{3}$ \citep[e.g.,][]{2019ApJ...871...12T} the minimum interval
between successive bursts would be $0.1 - 1$ s. In our model, lower twists would be associated with longer minimum recurrence times.

If magnetoelastic torsional oscillations follow the initial slippage event, as
observed in some magnetar short bursts, multiple FP dislocation events could
occur. The period of crustal torsional oscillations,
$1/\nu_{\rm osc} \sim (50-300 \rm \, Hz)^{-1}$ is generally shorter than the
twist charge depletion timescale
$\ell_B \kappa\rho_{\textrm{burst}}/j_{\textrm{twist}}$. However, for oscillations with large amplitude, the radio emission mechanism can operate for larger persistent twists. The cascade multiplicity dependency on the amplitude expected to be rather weak \citep[e.g.,][]{2019ApJ...871...12T}, hence, the larger current caused by larger twist would lead to faster clearing of flux tubes. Then, multiple nonstationary pair avalanches and radio bursts may transpire during such torsional oscillations. Since core-crust coupling is known to damp such
oscillations on a timescale of $\sim 0.2-2$ s
\citep{2006MNRAS.368L..35L,2014ApJ...793..129H,2019ApJ...871...95M}, the
duration and number of such time-clustered events ought be limited to a few
events in $\sim 2$ s time intervals, or up to when the oscillation amplitude
$\xi$ is too small to initiate pair cascades and satisfy
Eq.~(\ref{eq:rho_inequality}). Furthermore, because of such damping and charge
loading, the FRB pulse fluences may be lower for events spawned in oscillations
than the initial pulses triggered in conditions of higher charge
starvation. These expectations are in general agreement with short-waiting-time
events in Figure~\ref{deltaTOA}. {Speculatively, millisecond timescale substructures within longer bursts \citep[e.g.,][]{2019ApJ...876L..23H,2019Natur.566..235C} might also arise from plasma blobs spawned by crustal oscillations. }

High linear polarization of individual bursts can be naturally explained in the
framework of our model. There are two orthogonal eigenmodes of propagation in a
magnetized plasma, with one generally dominant
\citep[e.g.,][]{1977PASAu...3..120M,1979AuJPh..32...61M}. \citet{2019MNRAS.483..359L}
recently argued that in magnetar magnetospheres, the dominant X-mode can enter the
so-called ``adiabatic walking'' regime
\citep[e.g.,][]{1979ApJ...229..348C,2010MNRAS.403..569W} when propagation
induces high linear polarization and the PA traces the geometry of the inner
magnetosphere. They estimated the freeze-out radius
$r_{\rm fo}$, where radio emission finally decouples from plasma, preserving the
acquired linear polarization, for the corotation plasma density. Here we estimate $r_{\rm fo}$ for much higher plasma density required by the twisted magnetosphere. Adopting Eq.~(18) from
\citet{2019MNRAS.483..359L} and substituting expressions for the charge
density through Eq.~\eqref{eq:j_twist} and the threshold on the twist of magnetic
field lines Eq.~\eqref{eq:Phi_max}, for $\nu_{\rm em} \sim 1$ GHz radio waves, the freeze-out radius is
\begin{equation}
\frac{r_{\rm fo}}{R_*} \lesssim  18 \, B_{0,14}^{1/3} \, R_{B,7}^{1/3} \, a_{0,5}^{-1/3} \, \nu_{\rm em, GHz}^{-1/3} \,  \sigma_{-3}^{1/3}\,\nu_{\mathrm{osc,\, kHz}}^{1/3}
\label{adiabat}
\end{equation}
where $R_{B} \sim 10^7 R_{B,7}$ cm is the radius of curvature of the field
lines, $a_{0} \sim 10^5 a_{0,5} \gg 1$ is the characteristic nonlinearity
parameter, which may be regarded as the induced electron Lorentz factor
$a_0 \sim e E_R/ (\omega m_e c) \lesssim 10^5 - 10^7$ by the high-intensity
radio pulse of angular frequency $\omega$. Here
$E_R \sim \sqrt{{\cal E}_{\rm iso}/R_*^3}$ is the characteristic electric field
of the radio pulse. From Eq.~\eqref{adiabat} it is clear that adiabaticity may be attained in large
zones of the magnetospheres. The freeze-out radius is generally larger than the radius
above which 1~GHz radio emission is transparent, $r_{\mathrm{em}}\gtrsim 7 R_*$,
estimated above, so that the radio waves can acquire high linear polarization
prior to vacuum propagation.

Periodicity in the PA variation ought to be a viable check for the model,
particularly in repeating FRBs with high linear polarization. In the canonical
rotating vector model \citep[RVM, ][]{1969ApL.....3..225R} for a static dipole,
for viewing angle $\zeta \in (0, \pi)$ and magnetic obliquity
$\alpha \in (0 ,\pi/2)$ with respect to the spin axis, the allowed parameter
space for which the PA has bounded $< 180^\circ$ variation is
$0 < \beta/2 < \pi/2 - \alpha$ where $\beta = \zeta - \alpha$ is the impact
parameter. Under these assumptions for RVM, it may be shown that
$ 2\chi^\prime\equiv \rm PA_{\rm max} - PA_{\rm min}$ is restricted to
$\chi^\prime \geq \alpha$. Hence, in our model $\alpha \lesssim 40^\circ$ (see
\S\ref{obsPA}); this result obtained under the assumption of no preferential
sampling of spin phases in bursts, is in contrast with the lighthouse effect in
pulsars. If there exist spin phases where radio emission is either unobservable
or not amenable to the coherent radio process, then gaps in the folded PA sweeps
may manifest; however, such gaps would imprint periodicity in arrival
times. The null-detection of periodicity in arrival times of FRB 121102 pulses suggests such beaming selection effects may be small/inconsequential and the pulses ought to sample any spin phase of the rotator, similar to magnetar short bursts.

\section{Summary and Outlook}
\label{discussion}

In this paper, motivated by the remarkable similarity between statistics of magnetar short bursts and FRB 121102, we suggest that some FRBs originate from magnetars with low magnetospheric twist. Short bursts in such magnetars would give rise to pulsar-like radio emission mechanisms along closed field lines linked to the active region powering the burst. The crucial component of our model is that the plasma density above the active region powering the magnetar burst is insufficient to screen the accelerating electric field induced by the dislocation of magnetic FPs following crustal slippage events. Moreover, in magnetars with high twist, plasma density in the closed field line zone would be too high to allow $\sim1$~GHz radio transparency from most of the closed field line region. Hence, for self-consistency, this mechanism can operate only in magnetars which cannot support high plasma densities in the closed field line zone; this sets an upper bound on the global twist. Such an object could be an aged magnetar which lost most of its twist by the decay of internal toroidal fields, a high-B pulsar undergoing magnetar-like activity or a younger magnetar in a mode of low twist.  The low twist can account for the absence of FRBs from Galactic magnetars, which are believed to have larger twists than the limit in this work, not only because of charge starvation during crustal dislocations, but also because of absorption of radio pulses in the closed zone. Based on the empirical data about pulsar radio emission efficiency, we assume a $10^{-4} -10^{-2}$ fraction of the calorimetric short burst energy can be released in form of FRB, which is ample to account for observed fluences in FRB 121102. 

{The proposed mechanism might not work well for magnetar giant flares \citep[which is consistent with nondetection of radio bursts in the 2004 giant flare of SGR 1806--20,][]{2016ApJ...827...59T}. The energy release in giant flares is much larger ($\sim10^{44}-10^{46}$ erg) and the spectral character is distinct (spectra extending to much higher photon energies) from short bursts \citep[e.g.,][]{2006csxs.book..547W}. Giant flares should arise from a qualitatively different physical origin than short bursts \citep[e.g.,][]{2001ApJ...561..980T,2016MNRAS.461..877V} possibly involving reconnection in a large  magnetospheric volume with large twists \citep[e.g.,][]{2012ApJ...754L..12P,2013ApJ...774...92P}. During giant flares, huge amounts of dense pair plasma is generated, but the pair cascades that produce this plasma may be quite different from those that might lead to coherent radio emission. Pair formation may be distributed over a large volume, and the leading process may be two-photon pair creation, i.e. such cascades might not involve fast screening of large sustained (on FRB timescales) electric fields in charge starved regions by newly generated plasma, that are presumed to be at the core of pulsar-like emission mechanisms. So, even at the very onset of a giant flare, conditions might be unfavorable for the generation of coherent radio bursts. Moreover, once dense plasma and photon fields are generated, they will suppress any further production of nonthermal particle populations and/or will be opaque for radio emission. }

The event rate of cosmological FRBs will clearly depend on the operating longevity of the progenitor. If the low-twist FRB mode is the next stage in the life of a typical magnetar, then the absence of FRBs in the Galactic magnetar population sets a lower bound on the age of FRB progenitors to be {$\gtrsim 3-10$ kyr \citep[SNR ages, cf.][]{2019MNRAS.487.1426B}}, though such a mode may not be long-lived (or prolific), given that field decay also may act on similar timescales and predicted crustal event rates decline strongly with age \citep[e.g.,][]{2013MNRAS.434..123V,2019MNRAS.487.1426B}. Yet, the progenitor may be a young magnetar in a state of low twist. Magnetars with large-scale twists might temporarily lose their twist on a timescale of $\sim 10^2-10^3$ days \citep[e.g.,][]{2017ApJ...847...85Y,2018MNRAS.474..961C},  suggesting that the FRB mode with low twist may be a substantial fraction of an active magnetar's lifespan. Models of large-scale slow untwisting in magnetars \citep{2009ApJ...703.1044B,2013ApJ...777..114B,2017ApJ...844..133C} predict a significant colatitudinal dependence to the local twist, with equatorial FPs less twisted than polar ones, and the low-twist equatorial cavity expanding with time. Then, small dispersive delays, secularly decreasing at the untwisting timescale but dependent on rotational phase, may be imprinted on pulses.

In Galactic magnetars, the timescale of damping of crustal oscillations due to core-crust coupling has been inferred to be $\sim 0.2-2$ s \citep{2014ApJ...793..129H,2019ApJ...871...95M}. This, along with the time to clear a flux tube of charges, limits the number of potential FRB recurrences in $\sim 0.2-2$ s time intervals, if associated with a single active region. In such short-waiting-time event clusters {(or within substructures of longer bursts)}, quasiperiodicity associated with the crustal torsional oscillations may become apparent in arrival times for large samples. Scrutiny of the time variation in the PA of bursts, particularly those with high linear polarization, could be also pivotal. A periodicity of order $\sim 1-10$ s in the PA variation, tracing the magnetic field structure as in the RVM, will be a ``smoking gun" of the pair-starved magnetar model \citep[also see][]{2019MNRAS.483..359L}. 

The high-energy nondetection by \cite{2017ApJ...846...80S} for FRB 121102 with a burst energy limit
of $\lesssim 10^{45}-10^{47}$ erg is consistent with the short burst picture (since
short burst energies are smaller by a factor $\ll10^{-2}$).  Photon splitting
and magnetic pair production in the magnetosphere will suppress signals above a
few MeV \citep{2019MNRAS.tmp..982H,2019arXiv190305648W}, suggesting lower energy observations would
be more promising. Time coincidence \textit{Fermi}-GBM {and \textit{Gehrels/Swift}-BAT} scrutiny of future nearby
FRBs might provide a stringent test of the model.

\acknowledgements

We thank Matthew G. Baring, Alice K. Harding, Jason Hessels, Demos Kazanas, Chryssa Kouveliotou
and George Younes and for helpful discussions and valuable feedback on this
manuscript. {We also thank the anonymous referee for constructive feedback.} ZW is supported by the NASA postdoctoral program. AT is supported by
the NSF grant 1616632 and \textsl{Chandra} Guest Investigator program
TM8-19005. This work has made use of the NASA Astrophysics Data System.

\bibliographystyle{aasjournal}
\bibliography{magnetarrefs}

\end{document}